\documentclass[aip,apl,reprint]{revtex4-1}
\usepackage{graphics}
\usepackage{float}
\usepackage{epsfig}
\usepackage{dcolumn}
\usepackage{amsmath}
\usepackage{multirow}

\begin{document}

\title{Generation and control of spin-polarized photocurrents in GaMnAs heterostructures}

\author{Anibal T. Bezerra}
\email{anibal@df.ufscar.br}
\affiliation{Departamento de F\' \i sica, Universidade Federal de S\~ ao Carlos,
13565-905, S\~ ao Carlos, SP, Brazil}
\affiliation{DISSE - Instituto Nacional de Ci\^ encia e Tecnologia de Nanodispositivos
Semicondutores, CNPq/MCT, Brazil}
\author{Leonardo K. Castelano}
\affiliation{Departamento de F\' \i sica, Universidade Federal de S\~ ao Carlos,
13565-905, S\~ ao Carlos, SP, Brazil}
\author{Marcos H. Degani}
\affiliation{Faculdade
de Ci\^ encias Aplicadas, Universidade Estadual de Campinas,
13484-350, Limeira, SP, Brazil}
\affiliation{DISSE - Instituto Nacional de Ci\^ encia e Tecnologia de Nanodispositivos
Semicondutores, CNPq/MCT, Brazil}
\author{Marcelo Z. Maialle}
\affiliation{Faculdade
de Ci\^ encias Aplicadas, Universidade Estadual de Campinas,
13484-350, Limeira, SP, Brazil}
\affiliation{DISSE - Instituto Nacional de Ci\^ encia e Tecnologia de Nanodispositivos
Semicondutores, CNPq/MCT, Brazil}
\author{Paulo F. Farinas}
\affiliation{Departamento de F\' \i sica, Universidade Federal de S\~ ao Carlos,
13565-905, S\~ ao Carlos, SP, Brazil}
\affiliation{DISSE - Instituto Nacional de Ci\^ encia e Tecnologia de Nanodispositivos
Semicondutores, CNPq/MCT, Brazil}
\author{Nelson Studart}
\affiliation{Departamento de F\' \i sica, Universidade Federal de S\~ ao Carlos,
13565-905, S\~ ao Carlos, SP, Brazil}
\affiliation{DISSE - Instituto Nacional de Ci\^ encia e Tecnologia de Nanodispositivos
Semicondutores, CNPq/MCT, Brazil}

\date{\today}

\begin{abstract}
Photocurrents are calculated for a specially designed GaMnAs semiconductor heterostructure. The results  reveal  regions in the infrared range of the energy spectrum in which the proposed structure is remarkably spin-selective. For such photon energies, the generated photocurrents are  strongly spin-polarized. Application of a relatively small static bias in the growth direction of the structure  is predicted to efficiently reverse the spin-polarization for some photon energies. This behavior suggests the possibility of conveniently simple switching mechanisms. The physics underlying the results is studied and understood in terms of the spin-dependent  properties emerging from the particular potential profile of the structure.
\end{abstract}

\maketitle

Generation and control of spin-polarized carriers are two basic ingredients required for a spintronic device. Systems yielding these ingredients have been intensively studied in the last few years.~\cite{zut04,Aws13,War13,Yu13,cas10} In particular, diluted magnetic semiconductors (DMS) have been under renewed attention.~\cite{chi99,Diet10,Mun12,Pasc12,Gra12,Sam12,Fuj13} The revival is due in part to the evidence of the existence of ferromagnetism observed in DMS with a consistent increase of the Curie temperature.~\cite{Ohta13,RUNG13,ber11,cha12} Recently, a study on laser-assisted spin-polarized current in DMS heterostructures showing the mechanisms that produce the spin-polarization and the photocurrents induced by THz radiation has been reported.\cite{olb12} The findings show a behavior of the photocurrent with an external magnetic field that evidences the spin-polarization of the photocurrent.

\begin{figure}
\linespread{1.0}
\includegraphics[width=6cm]{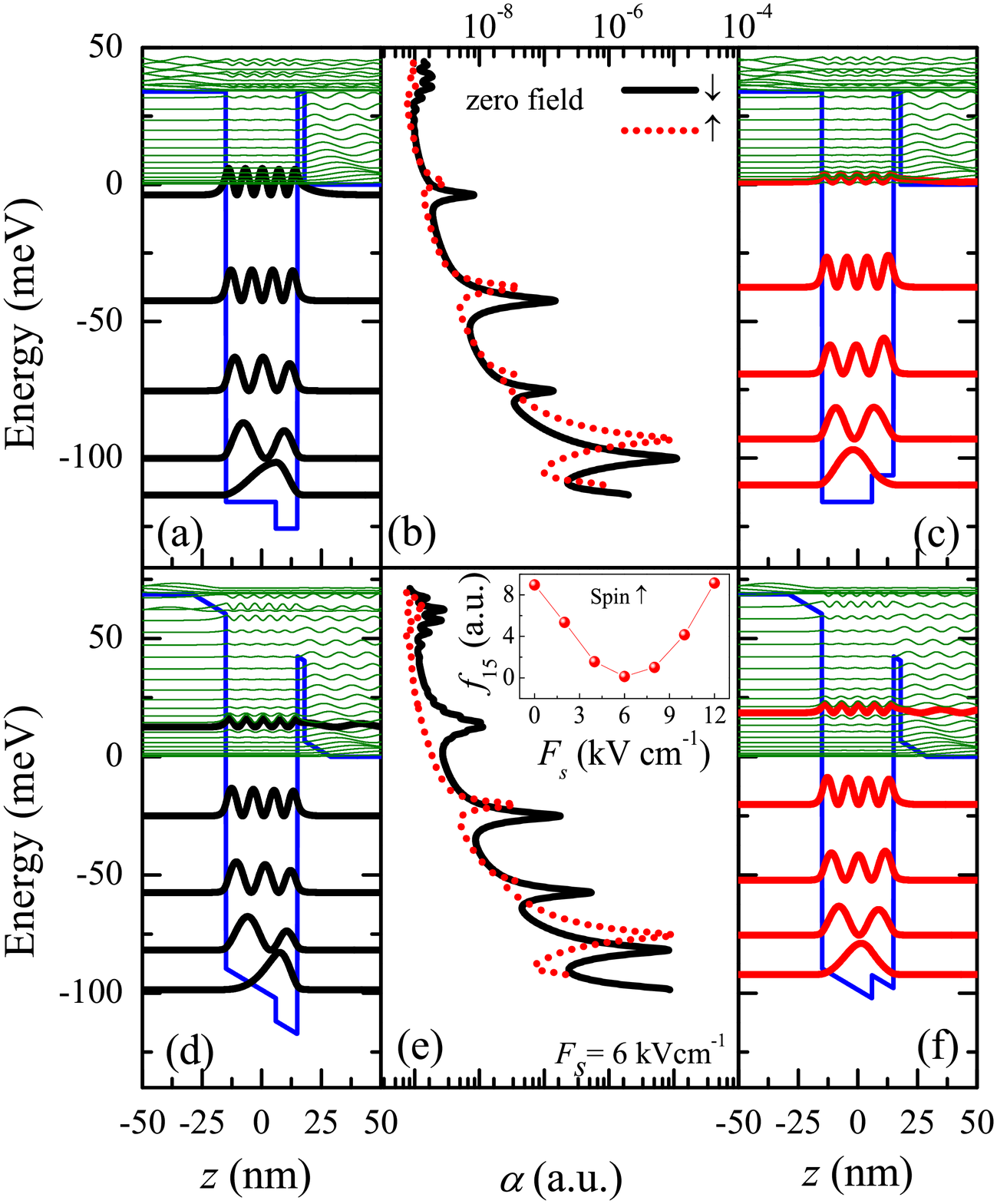}
\caption{\label{figu1} (a) Spin $\downarrow$ potential profile of the structure for zero electric field, showing the modulus squared wave functions. (b) Absorption spectrum for the spin $\downarrow$ (solid black) and $\uparrow$ (dotted-red) for zero field.  (c) Same as (a) for the spin $\uparrow$. (d) Same as (a) for $F_s=$ 6 kVcm$^{-1}$. (e) Same as (b) for $F_s=$ 6 kVcm$^{-1}$. (f) Same as (c) for $F_s=$ 6 kVcm$^{-1}$. The inset on the panel (e) presents the oscillator strength of the transition between the spin $\uparrow$ ground state (E$_{1}$) and the fifth state (E$_{5}$), showing a pronounced decrease for electric fields around 6 kVcm$^{-1}$.}
\end{figure}
 
Improvements of the growth techniques have been increasing the potential for new designs based on III-V DMS, like GaMnAs, for spin devices. Some advances, like the significant increase of maximum Mn concentration which can be incorporated substitutionally,~\cite{WAN13, DOL12, CHAP13} and the understanding of GaAs-related structures give motivation for new studies on GaMnAs structures. 

In this work we propose a GaMnAs-based DMS nanostructure that enables the generation of a spin-polarized photocurrent and its control through a static electric field (that we call ``bias'') applied in the growth direction. We investigate this system through the direct analysis of the time-dependent solution of the Schr\"odinger equation for the potential profile's design in the presence of an oscillating electric field playing the role of excitation photons shed on the system. The photocurrent is calculated from the full time-dependent solution and from the imaginary time solution we are able to compute ground and excited states' properties, like energies and couplings. Thanks to the special design of the quantum-well, varying the bias yields control the oscillator strengths, directly affecting the absorption probabilities. In addition, a design-generated spin selectivity due to different potential profiles for different spins, results in spin-dependent couplings, leading to a spin dependent response to the external bias, thus allowing the control of the photocurrent's spin polarization.
 
The potential profile of the proposed structure is depicted in Fig~\ref{figu1}. It consists of a 30 nm thick quantum well formed by two layers: a 21 nm thick GaAs layer, adjacent to a 9 nm thick Ga$_{0.91}$Mn$_{0.09}$As layer. On the left-hand side of the well, we include a wider barrier of Al$_{0.19}$Ga$_{0.81}$As which is intended to block the electron flux towards the left region, that can be different from zero even in the absence of static biases. On the right-hand side of the well, we include a 3 nm Al$_{0.19}$Ga$_{0.81}$As narrow barrier, followed by a Al$_{0.15}$Ga$_{0.85}$As wide layer.~\cite{VUR01} When a magnetic field is applied along the axis perpendicular to the layers, it causes an energy splitting in the DMS due to the well known Giant Zeeman Splitting.~\cite{SAF09} This is known to be caused by a hole-mediated \textit{sp-d} exchange interaction between electrons and the local moments of the open \textit{d}-shells in the Mn atoms,~\cite{SAF09,MIN11} and originates a spin-dependent effective potential.~\cite{MIT07}

The photocurrent of a single electron (placed initially in the ground state of this effective potential) is evaluated by a numerical approach that  obtains the full time depend state from the time-evolution operator  $e^{-iH\Delta t /\hbar}$,~\cite{DEG11,DEG10} where $\Delta t$ is a small time increment and $H$ is the systems Hamiltonian, given by
\begin{equation}
H = -\frac{\hbar^{2}}{2m^{*}}\frac{d^{2}}{dz^{2}}+ V_{\sigma}(z) - ez(F_{s} - F_{d}\sin\omega t) \\ +H_{\parallel}(x,y),
\end{equation}
where $m^{*}$ is the electron effective mass, $F_{s}$ is a static electric field produced by an external bias along $z$, and $F_{d}$ is the amplitude of a time-oscillating electric field with frequency $\omega$. In what follows $\hbar\omega$ will be referred to as the ``photon energy'', $F_d$ as the ``photon field'' amplitude and $F_{s}$ as the electric field amplitude. $H_{\parallel}(x,y)$ is assumed to be translationally invariant within the \textit{x},\textit{y} plane thus its quantum numbers (which are Landau level indexes in case) are conserved during the generation of the photocurrent. $V_{\sigma}(z)$ is the spin-dependent potential given by
\begin{equation}\label{EQ02}
V_{\sigma}(z) = V(z) + \frac{5}{3} N_{0}\alpha\sigma_{z}x_{eff}JB_{J}\left(\frac{2\mu_{B}B}{k_{B}(T+T_{0})}\right),
\end{equation}
where $B=$ 5 T is the applied magnetic field,  $V(z)$ is the bare structure's potential profile along z, $x_{eff}=0.018$ is the effective Mn concentration, $J=\frac{5}{2}$ the spin of the Mn ions, $\sigma_{z}=\pm\frac{1}{2}$ corresponding to spin-up ($\sigma=\uparrow$) and -down ($\sigma=\downarrow$) components, respectively. $N_{0}\alpha=-0.27$ eV is the $sp-d$ exchange constant, $B_{J}$ is the Brillouin function, $\mu_{B}$ is the Bohr magneton, $k_B$ is the Boltzmann constant and $T_{0}=12.5$ K accounts for the Mn-Mn antiferromagnetic interaction.~\cite{GAN09,MIT07}

Time evolution is achieved by using the split operator method~\cite{DEG10} 
\begin{equation}
e^{-i(T+U)\Delta t/\hbar} = e^{-iU\Delta t/2\hbar}e^{-iT\Delta t/\hbar}e^{-iU\Delta t/2\hbar}+O(\Delta t^{3}),\nonumber
\end{equation}
where $T$ and $U$ are the kinetic and potential energy operators, respectively. Successive applications of the exponential operator brings the $t = 0$ wave function to the time dependent wave function. Performing the time-evolution in imaginary time for $F_{d}=0$,  yields the eigenstates and eigenergies of the static potential.~\cite{DEG10}
With this numerical method, external static and dynamic fields are treated exactly with arbitrary precision.

The photocurrent is obtained from the current density
\begin{equation}
J_{\sigma}(\hbar\omega,t)=\left.\Re\left[\frac{\hbar}{im^{*}}\Psi_\sigma^{*}(\hbar\omega,z,t)\frac{\partial\Psi_\sigma(\hbar\omega,z,t)}{\partial z}\right]\right|_{z_R},
\end{equation}
where $\Psi_\sigma(\hbar\omega,z,t)$ is the time-evolved wave function for the component $\sigma$ and $z_R=45$ nm is a position taken far outside the confining region. The current density $J_\sigma(\hbar\omega,t)$ is integrated in time to give the spin-dependent photocurrent,
\begin{equation}\label{EQ04}
I_\sigma(\hbar\omega)=\frac{e}{T_{p}}\int^{T_{p}}_{0}J_\sigma(\hbar\omega,t)dt,
\end{equation}
where $T_{p}$ is the duration of an oscillating electric field pulse,~\cite{MAI11} here chosen as 3 ps. 

Another useful quantity that can be extracted directly from the split-operator method, using the imaginary time evolution, is the absorption spectrum, which is calculated through
\begin{equation}\label{EQ06}
\alpha(E) = \sum_{n}\frac{f_{1n}\Gamma}{2\pi\left[(E-E_n)^{2}+ \frac{1}{4}\Gamma^{2} \right]},
\end{equation}
where $f_{1n}$ is the oscillator strength between the ground state and the $n^{th}$ excited state, defined as
\begin{equation}\label{EQ07}
f_{1n} = \frac{2m^{*}}{\hbar^{2}}\Delta E_{n}|\left\langle\psi_{n}|x|\psi_{1}\right\rangle|^{2},
\end{equation}
with $\Delta E_n = E_n-E_{1}$ being the energy separation between the $n^{th}$ state $\psi_{n}$ and the ground-state $\psi_{1}$ wave functions, whose moduli are depicted in Fig.~\ref{figu1} (panels a, c, d, and f). $\Gamma=$ 2 meV has been adjusted to resolve the peaks in the curves of the absorption spectra.

We present in Fig.~\ref{figu1} potential profiles of the spin $\downarrow$ and $\uparrow$ components for zero electric field and $F_s=6$ kVcm$^{-1}$. The width and depth of the QW have been engineered to yield the following differences for spins $\uparrow$ and $\downarrow$. For the spin $\downarrow$ component, there are four bound states in the well with a fifth state E$^{\downarrow}_{5}$ which can be either bound or continuum state for different biases, as shown in the Figs.~\ref{figu1}(a) and (d). For the spin $\uparrow$ component, we have four bound states, however the fifth state E$^{\uparrow}_{5}$ is in the continuum independent on the bias (Figs.~\ref{figu1}(c) and (f)). Thus, the structure is engineered to work as a spin-dependent quantum infrared detector (QWIP),~\cite{DEG11} for which the photocurrent's spin polarization can be controlled by a static bias. 

In order to explore these features, it is worth investigating numerically the controll values, and the physical origin of the transitions involved. Whence, before looking at the photocurrent obtained in our simulations, we turn to the absorption spectrum shown in Fig.~\ref{figu1}. Fig.~\ref{figu1}(b) shows the absorption for both spin-components, with the peaks aligned with the final state of the respective transition, and the results have been obtained using the ground state as the initially occupied state. The peaks correspond to the transitions between ground and excited states for spin $\downarrow$ (solid black curve) and spin $\uparrow$ (dotted red curve) components. In special we can observe the zero-energy peaks which correspond to the absorption between the ground state and the E$^{\downarrow,\uparrow}_{5}$ states, majorly responsible for the generation of current. For zero bias, although this specific transition is allowed for the spin $\downarrow$ component, the final state is bound and localized in the well (see Fig.~\ref{figu1}(a)), thus no photocurrent is expected for the spin $\downarrow$. We only expect generation of photocurrent for this state via multi-photon excitation processes, which are reduced when the dynamic field intensity is small ($F_d=1$ kVcm$^{-1}$).~\cite{DEG11} The transition for the spin $\uparrow$ component is also allowed, but the latter is an extended state in the continuum for any bias. Therefore, for zero bias we expect to see only spin $\uparrow$ photocurrent.

As the static bias is increased from $F_s=0$ to $F_s=6$ kVcm$^{-1}$, the E$^{\downarrow}_{5}$ state is dragged to the continuum as shown in Fig.~\ref{figu1}(d), and becomes extended. The transition between spin $\downarrow$ ground state and the E$^{\downarrow}_{5}$ state remains allowed, as shown by the persistence of a peak around 10 meV in the spin $\downarrow$ absorption spectrum (full black curve in Fig.~\ref{figu1}(e)). Thus, changing the static field from $F_s=0$ to to $F_s=6$ kVcm$^{-1}$, will lead to the generation of spin $\downarrow$ photocurrent. The spin $\uparrow$ component is, however, expected to be reduced for the following reasons. 

Although the E$^{\uparrow}_{5}$ state remains in the continuum for any value of the applied bias, its coupling to the ground state is reduced, as shown in the inset of the Fig.~\ref{figu1}(e). We see that the oscillator strength of the transition between the spin $\uparrow$ ground state and the E$^{\downarrow}_{5}$ state, $f_{15}$, is clearly dependent on the electric field, showing a minimum for $F_s=6$ kVcm$^{-1}$. This dependence of the oscillator strength on the static electric field is related to the QW structural asymmetry, imposed by keeping the DMS layer only in a limited region in the QW, and the effect this has upon the spin $\uparrow$ ground state. Due the relatively large potential hights involved, the lower energy eigenfunctions may be seen as deviations from states of a symmetrical infinite quantum well, and hence they guard an approximate parity character in \textit{z}, about the center of the well even though the potential has no parity defined in the z direction. By adjusting the bias this ``parity'' can be tuned in order to suppress (or reinforce) specific couplings, as shown in Fig.~\ref{figu1}(f). The matrix element $\left\langle\psi_{n}|z|\psi_{1}\right\rangle$ contains states that become nearly even in \textit{z} as the bias is increased from $F_s=0$ to 6 kVcm$^{-1}$, and the oscillator strength is drastically reduced for the spin $\uparrow$ component as the bias is increased, with the suppression of the transition. As we can observe, for zero field, these reduced oscillator strength is already present for the spin $\uparrow$ component, as shown by the reduced peaks in the absorption (Fig.~\ref{figu1}(b)). For such a bias, the spin $\uparrow$ ground state not fully resembles the whole QW ground state and the absorption of the even-like states in not totally suppressed. Further increasing the electric field, the spin $\uparrow$ wave functions Stark shift to the right-hand side of the QW broking the even-like condition and the oscillator strength is increased (see the inset in the Fig.~\ref{figu1}(e)). The spin $\downarrow$ ground state, independently of bias, stays localized in the right-hand side of the structure, see the Figs.~\ref{figu1}(a) and (d). Therefore, the spin $\downarrow$ oscillator strength is not directly affected by changes in the electric field. 

This interesting behavior of the oscillator strength is highlighted in the absorption spectrum of the spin $\uparrow$ component for $F_s=$ 6 kV cm$^{-1}$ (dotted red curve in Fig.~\ref{figu1}(e)). The transition between the spin $\uparrow$ ground state and the E$^{\uparrow}_{5}$ is quenched for such a bias. In this way, being the transition forbidden, even matching the energy separation between the spin $\uparrow$ ground state and the E$^{\uparrow}_{5}$ state, we do not expect to generate spin $\uparrow$ photocurrent for $F_s=$ 6 kV cm$^{-1}$.

\begin{figure}
\linespread{1.0}
\includegraphics[width=6.0cm]{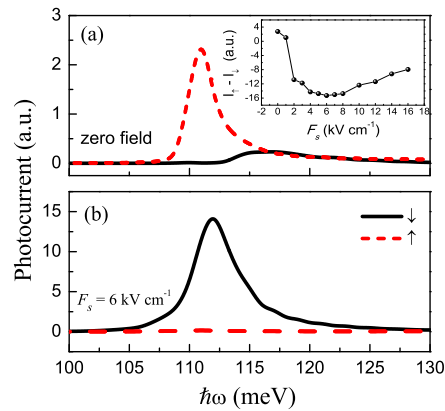}
\caption{\label{figu3} (Color online) Spin $\downarrow$ (solid black) and $\uparrow$ (dashed red) photoccurent signal as a function of the energy of the exciting field for (a) zero field, (b) $F_s=$ 6 kV cm$^{-1}$. The inset shows the intensity differences between the spin $\uparrow$ and $\downarrow$ peaks of photocurrent, as a function of the electric field.}
\end{figure}

Finally, the photocurrent responses are shown in Figs.~\ref{figu3}(a) and (b), for zero field, and 6 kV cm$^{-1}$, respectively. In what follows, the photocurrent was evaluated in a purely coherent fashion, without considering ohmic effects. We achieved spin-polarized photocurrent with a peak around 110 meV for both spin polarizations, switched by the applied electric field. As expected, for zero field (Fig.~\ref{figu3}(a)) we have predominance of the spin $\uparrow$ photocurrent, whose polarization was obtained by maintaining the E$^{\downarrow}_{5}$ bound and the E$^{\uparrow}_{5}$ in the continuum. Increasing the field to $F_s=$ 6 kV cm$^{-1}$, we have almost exclusively spin $\downarrow$ current (Fig.~\ref{figu3}(b)), thanks to decreasing of the spin $\uparrow$ oscillator strength, which quenches the absorption probability between the spin $\uparrow$ ground state and the E$^{\uparrow}_{5}$ excited state, reducing drastically the spin $\uparrow$ photocurrent. 

As noted, for the specific structural design presented here, it were showed only the optimal electric field conditions which generate the bias-selective spin polarization, for instance, zero field and $F_{s}=$ 6 kV cm$^{-1}$. The photocurrent behavior for other electric fields is summarized in the inset of the Fig~\ref{figu3}(a), which shows the difference between the intensity of the photocurrent peaks for the spin $\uparrow$ and $\downarrow$ components as a function of the electric field. As shown in the inset, for zero electric field, we achieved the maximum spin $\uparrow$ polarization. Increasing the bias, the E$^{\downarrow}_{5}$ state is dragged through the continuum increasing the spin $\downarrow$ photocurrent. The spin $\uparrow$ oscillator strength is decreased with increasing the electric field (inset inf Fig.~\ref{figu1}(e)), reducing the spin $\uparrow$ photocurrent. Thus, up to $F_{s} =$ 6 kV cm$^{-1}$, we have an intermediary scenario, presenting both spin polarizations. After this field, the spin $\uparrow$ oscillator strength is increased restoring the condition with both spin polarizations. Moreover, for even higher bias, the remaining bound states can also be dragged to the continuum and contribute to the photocurrent. However this contribution occurs for different spectral domains.

In summary, we have shown a light-assisted generation and control of spin-polarized current in a DMS heterostructure. We proposed an asymmetric two-layered QW structure with a GaMnAs layer on the well. The spin polarization of the photocurrent was achieved by dealing with the eigeinstates wave functions, handling the oscillator strength of the transition between the spin-dependent ground states and the excited states. As observed, the proposed structure is sensitive to photons in the infrared ($\sim$11 $\mu$m) region, offering the possibility of working as spin-sensitive infrared quantum well photodetectors. Notwithstanding, the method employed to control the spin-polarization of the photocurrent in GaMnAs materials can be also extended to other III-V family materials as the nitrates GaMnN, as well to the II-VI family materials as the ZnMnSe and CdMnTe, opening the possibility of different energy ranges of operation. 

This work is funded by DISSE-Instituto Nacional de Ci\^encia e Tecnologia de Nanodispositivos Semicondutores and Conselho Nacional de Desenvolvimento Cient\'ifico e Tecnol\'ogico (CNPq). MHD and MZM acknowledge financial support from Funda\c{c}\~ao  Amparo \`a Pesquisa do Estado de S\~ao Paulo (FAPESP). LKC acknowledges support from FAPESP under grant No. 12/13052-6.

\linespread{1}

\end{document}